\documentclass[conference]{IEEEtran}
\IEEEoverridecommandlockouts

\usepackage{times}            
\usepackage[english]{babel}   
\usepackage[ansinew, utf8]{inputenc}
\usepackage[T1]{fontenc}      
\usepackage[sort&compress,numbers]{natbib}	
\usepackage{amsmath,amssymb}
\usepackage{graphicx}
\usepackage[colorlinks=false,pdfborder={0 0 0}]{hyperref}
\usepackage{units}

\usepackage{xcolor}
\usepackage{booktabs}
\usepackage{cleveref}

\usepackage[acronym, shortcuts, nohypertypes={acronym}]{glossaries}
\newacronym{MAE}{MAE}{mean absolute error}
\newacronym{PCC}{PCC}{Pearson correlation coefficient}
\newacronym{MSE}{MSE}{mean squared error}

\begin{document}

\title{Audio-based Step-count Estimation for Running -- Windowing and Neural Network Baselines\\
\thanks{This work was funded from the DFG’s Reinhart Koselleck project No. 442218748 (AUDI0NOMOUS) and the Zentrales Innovationsprogramm Mittelstand (ZIM) grant agreement No. 16KN069455 (KIRun).}
}


\author{\IEEEauthorblockN{
Philipp Wagner\IEEEauthorrefmark{1}\IEEEauthorrefmark{2},
Andreas Triantafyllopoulos\IEEEauthorrefmark{1}\IEEEauthorrefmark{2},
Alexander Gebhard\IEEEauthorrefmark{1}\IEEEauthorrefmark{2},
Bj\"orn Schuller\IEEEauthorrefmark{1}\IEEEauthorrefmark{2}\IEEEauthorrefmark{3},
}
\IEEEauthorblockA{\IEEEauthorrefmark{1}CHI -- Chair of Health Informatics, Technical University of Munich, MRI, Munich, Germany}
\IEEEauthorblockA{\IEEEauthorrefmark{2}EIHW -- Chair of Embedded Intelligence for Health Care and Wellbeing, Augsburg, Germany}
\IEEEauthorblockA{\IEEEauthorrefmark{3}GLAM -- Group on Language, Audio, \& Music, Imperial College London, UK}
\IEEEauthorblockA{Email: andreas.triantafyllopoulos@tum.de}
}



\maketitle

\begin{abstract}
In recent decades, running has become an increasingly popular pastime activity due to its accessibility, ease of practice, and anticipated health benefits. 
However, the risk of running-related injuries is substantial for runners of different experience levels.
Several common forms of injuries result from overuse -- extending beyond the recommended running time and intensity.
Recently, audio-based tracking has emerged as yet another modality for monitoring running behaviour and performance, with previous studies largely concentrating on predicting runner fatigue.
In this work, we investigate audio-based step count estimation during outdoor running, achieving a mean absolute error of 1.098 in window-based step-count differences
and a Pearson correlation coefficient of 0.479 when predicting the number of steps in a 5-second window of audio.
Our work thus showcases the feasibility of audio-based monitoring for estimating important physiological variables and lays the foundations for further utilising audio sensors for a more thorough characterisation of runner behaviour.
\end{abstract}


\section{Introduction}

Multiple demographic studies ascertain that running is one of the most popular leisure
activities across Europe and the United States \cite{running0, running3, running4, running5}.
According to Pereira et al.\ \cite{running4}, the prevalence of running varies between 12 to 34\,\% in Europe and 15\,\% in the USA
and Australia, with a significantly rising trend in the past decades \cite{running0, running3, running4, running1}.
While outdoor running entails a multitude of health and wellbeing benefits, the risk of injuries is relatively high, especially in combination with overuse, mileage, or previous injuries \cite{running0, running3}.
This is especially true for newcomers to the sport which may not have the required experience to properly adapt their training.
For this reason, the use of automatic monitoring technologies that can track running behaviour over time has been attracting increasing interest from the research community.

The majority of previous research has focused on wearable sensors which monitor different physiological variables.
For example, continuously tracking biomechanical signals can offer a lot of information that can be applied for injury prevention.
To that end, Schütte et al.\ \cite{running6} examined the effects of fatigue on body stability during highly demanding running sessions. 
In their experiments, running data was captured and tracked with chronometers, heartbeat sensors, and accelerometers. 
The results demonstrated that changes in acceleration and step frequencies can negatively affect the body’s stability under fatigued circumstances, leading to a higher risk of injury. 
In similar fashion, Milner \cite{running8} and Bowser et al.\ \cite{running9} were able to associate the origin of tibial stress fractures, which account for about 50\,\% of running injuries, with changes in impact loading and ground reaction forces happening at the occurrences of step motions.
Such works showcase the potential upside of using sensor-based monitoring for injury prevention.
However, their findings are primarily based on specialised equipment (e.\,g.\ wristbands or heartrate sensors) which a user will have to acquire.
This presents a barrier to entry for newcomers that may wish to benefit from guided training without the additional overhead of buying new sensors.

Recent studies have shown that acoustic monitoring offers a relevant alternative to traditional sensors in replacing or complementing the information-gathering process. 
Pirscoveanu et al.\ \cite{running10} discovered that the peak amplitude values of step sounds were more pronounced for fatigued runners, presumably due to an increased vertical loading rate, which refers to the step impact forces.
The results of Triantafyllopoulos et al.\ \cite{kirun0} show that acoustic sensors can be used for fatigue prediction and achieve similar results as in non-acoustic experiments \cite{running11}.
Oliveira et al. \cite{running13} go a step further by building on the results of \cite{running10} in attempting to predict vertical ground reaction forces in running sounds, with the assumption of correlation to fatigue.
On a different note, Gebhard et al.\ \cite{kirun1, kirun2} explored the application of audio to detect and classify different types of running surfaces and heart rates, which allows a more holistic understanding of a training routine.

These prior works illustrate the potential of acoustic monitoring for running behaviour.
Our contribution tries to build on these results by exploring the feasibility of using neural networks to detect and count occurrences of step sounds.
We formulate our task as \emph{the regression of the number of steps within a fixed time window}.
To this end, we explore a variety of audio data processing techniques, models, and machine-learning procedures.
Our work can be seen as the crucial first step in an full-blown, automatic processing pipeline for running analysis based on audio.
Predicting the number of steps allows for the subsequent computation of different variables (e.\,g.\ speed or acceleration).
Additionally, it facilitates the future, more granular prediction of ground reaction forces and their connection to runner fatigue.

The remainder of this paper is structured as follows: 
In \cref{sec:dataset}, we outline the data used for this study.
\Cref{sec:setup} discusses our experimental setup and \cref{sec:results} presents our results, followed by our concluding remarks in \cref{sec:conclusion}.


\section{Dataset}
\label{sec:dataset}

The KIRun dataset \cite{kirun0} -- collected in our previous work -- contains 188 audio files of running recordings.
These recordings were generated by 51 runners with an average runner age of 40 years, ten years of running experience, and a gender distribution of seven to one in favor of female runners. The average running duration of all the files is rounded to 45 minutes.
There are three types of running sessions in the data:
a) indoor treadmill runs,
b) outdoor running in specified tracks,
c) free outdoor runs.
The running surfaces comprise different running scenarios, such as forest paths, asphalt roads, or treadmills.
Audio data was recorded by an armband-encased smartphone attached to each runner's hand.
Furthermore, linear and rotational acceleration data was collected using the SensoRun\texttrademark system which comprises two inertial measurement units (IMUs) (Bosch, BMI160) attached to each tibial head.
These acceleration values allowed for the exact specification of step timings, which constitute our ground truth.
The step annotations were initially divided into five splits of data $\{s0,s1,s2,s3,s4\}$.



\section{Experimental Setup}
\label{sec:setup}

As mentioned, we formulate our task as the prediction of the number of steps within a given time window using the audio recorded from the 
runners' 
smartphones.
The following sections describe our methodological approach for achieving this goal.



\subsection{Preprocessing}

A large part of our preprocessing stage revolved around the windowing of audio files to shorter segments of audio, in order to provide an efficient input shape for the training stage. For example, the majority of models were trained on Mel-spectrogram inputs of shape $(C, F, T) = (1, 64, 500)$, where C denotes the channels, F the frequency, and T the time axis..
The resulting short clips of audio represented the inputs for the subsequent steps of our pipeline.
We additionally experimented with (non-overlapping) window sizes of $\{5, 10, 20\}$ seconds.
Using the step timings identified through SensoRun, we computed the total number of steps within each window, which constitutes our prediction target.


\subsection{Modelling}

We experimented with a variety of different neural network architectures for the prediction of steps.
Our first aim was to explore different architecture `classes' to gauge their suitability for the task, before proceeding to optimise them.
We initially tested a simple feed-forward network with 5 fully-connected hidden layers of size 256, and the ReLU activation function after each, but the last layer.
This simple benchmark has the advantage that it accepts different feature representations as a single feature vector.
Following that, we continued with recurrent neural networks (RNNs) and their combination with convolutional layers (CRNNs).
We tested an LSTM and GRU cells with different settings regarding the number of layers, hidden size and bidirectionality and the CRNNs described in \cite{facrnn, fdycrnn}.
Finally, we utilised convolutional neural networks (CNNs) and audio-based Transformers derived from the Vision Transformer (ViT) \cite{vit}.
CNNs currently achieve state-of-the-art performances in audio-related tasks, and a wide range of models for tasks such as audio tagging or acoustic scene classification is publicly available \cite{panns}.

The contribution of Pretrained Audio Neural Networks (PANNs) \cite{panns}, for example, includes a variety of models based on the VGG, ResNet, or MobileNet architectures. These models are well-suited for our task and were initially designed for audio tagging on the AudioSet \cite{audioset} dataset.
Among the models we trained were Cnn6, Cnn10, Cnn14, ResNet54, MobileNetV1/V2; all require the same input-shape of Mel-spectrograms.
In particular, the Wavegram-CNN is a specialisation of the VGG architecture that accepts waveforms as input, which are transformed into a learnable time-frequency representation called \textit{Wavegram} as the result of multiple convolution and pooling operations.
A Wavegram-Logmel-CNN, such as the WvLmCnn14 that we tested, further combines the best of both worlds, by combining the Wavegram with Mel-spectrogram features.

A similar approach of basing the model design on pre-existing network architectures can be found in PSLA \cite{psla}. 
The main proposition of PSLA is an optimised training pipeline and a CNN based on EfficientNet \cite{efficientnet}.
Similarly, the Audio Spectrogram Transformer \cite{ast} and other Transformer-based models such as HTS-AT \cite{htsat} or PaSST \cite{passt} are one of the latest advancements in audio recognition.
AST is an adaption to the ViT, that specialises on spectrogram input. HTS-AT reduces computation cost by limiting the number of learnable weights, and PaSST uses \textit{patchout}, an integrated augmentation technique, to elevate its performance and efficiency.
Code and pre-trained weights are publicly available for all these models, which enabled their rapid prototyping and also allowed us to test the impact of transfer learning.


\subsection{Training Setup}

All models except Transformer-based ones were trained with the Adam optimiser, using an initial learning rate of $1^{-3}$. 
The Transformer models were trained with the \linebreak AdamW optimiser instead ($\beta_1 = 0.9, \beta_2 = 0.999, \epsilon = 1^{-7}, \linebreak \lambda = 0.01$), using an initial learning rate of $1^{-4}$.
A learning rate scheduler monitored changes in a target metric between 100 epochs of training. After five consecutive epochs without improvement, the learning rate was multiplied by 0.9. This reduction process was done either until a minimal prior selected learning rate was reached or the training diverged, meaning no improvements occurred for a substantial amount of time. A mini-batch size of 32 was used for all experiments. 

As described in \cref{sec:dataset}, the dataset is split in five folds, each with a training, validation, and testing partition roughly following a 60\%-20\%-20\% distribution.
These splits were designed for runner independence, i.\,e., the train set does not contain any runners that are also within the validation or test set of the same partition. 
Given a large number of experimental configurations (models, feature sets, hyperparameters), we conducted some preliminary experiments using just a single split (s0) to avoid overfitting.
This was used to gauge the performance of each architecture and prune down the number of configurations for the proper cross-validation setup.
This meant training the same model 5 times on each split with different unseen test datasets and taking the mean performance out of all five evaluations.
For all experiments, we used the \ac{MSE} loss, as it is well-suited for regression. Both train and validation losses were monitored to track whether the model can generalise or is prone to overfit the training data.
As metrics, we report the \ac{MAE} and the \ac{PCC}; the first can be useful in application where an exact step count is needed, and the second in application where only the trend is relevant (e.\,g. to measure relative speed).

\section{Results}
\label{sec:results}




Our preliminary results on s0 with a 5-second window are shown in \cref{t:archresults2}, where we rank different architectures with respect to \ac{MAE} (ascending).
In addition to our models, we include a naive baseline which always predicts the mean of the training set.
Overall CNNs were the best-performing models on this partition, in particular WvLmCnn14 and PSLA. The next closest result was achieved by the PaSST-S Transformer model. 
Relatively high values of correlation (CC) tell us that there is a moderate linear fit between the prediction results and label distribution for these models. Also, the \ac{MAE} values are improved from the naive baseline by up to 25\,\% for certain CNNs. The HTS Transformer model performed worse than the PaSST-S Transformer, which emphasises the success of the \textit{patchout} strategy.

\begin{table}[t]
\begin{center}
\caption{
Preliminary results on split s0.
Showing PCC and MAE results.
}\label{t:archresults2}
\scalebox{.9}{
\begin{tabular}{lcc}
    \hline
        Model & PCC & MAE\\
        \hline
        PSLA \cite{psla} & \textbf{0.534} & \textbf{0.814}\\
        WvLmCnn14 \cite{panns} & 0.528 & 0.847\\ 
        PaSST-S \cite{passt} & 0.419 & 0.935\\
        FNN & 0.000 & 0.946 \\
        HTS-AT \cite{htsat} & 0.210 & 0.959\\
        Naive Baseline:					& - & 1.100\\ 
    \hline
\end{tabular}}
\end{center}
\end{table}

\subsection{Cross Validation}

Following the preliminary results on s0, we proceed to benchmark the three best-performing models of different neural architectures on the full cross-validation setup, shown in \cref{t:bestresults}. 
All models were again trained for 100 epochs on the remaining four splits. 
By taking the mean of all test evaluation metrics across each partition, we received a more robust outcome that considers more combinations of runner files with distinguishable running sound properties. 
Results showed that PSLA achieved the lowest \ac{MAE}, WvLmCnn14 the highest correlation and lowest MAE, and the PaSST-S Transformer fell behind the CNNs by a small margin.



\begin{table}[t]
\begin{center}
\caption{
5-fold cross validation
for the best-performing models.
}\label{t:bestresults}
\scalebox{1}{
\begin{tabular}{c|cc}
\hline
 Model       & PCC      &MAE \\
\hline
PSLA \cite{psla}        & 0.442 & \textbf{1.098}\\
WvLmCnn14 \cite{panns}  & \textbf{0.479} & 1.171\\
PaSST-S  \cite{passt}   & 0.443 & 1.186\\     
Naive Baseline			& - & 1.421\\
\hline
\end{tabular}}
\end{center}
\end{table}

\subsection{Ablation Studies}

Having seen the general learning behaviour of networks for the regression task, we subsequently evaluated the alteration of hyper-and meta-parameters. 
Among the modifications were different input features, annotation usage, and the impact of transfer learning. 
The approach and challenge of this stage were focused on maintaining comparability between neural models of different core architectures. 
The experiments were again conducted on split s0 of the provided splits.

\subsubsection{Data Augmentation}
We further compared the results of three different data augmentation methods, which are commonly used for audio recognition: SpecAugment \cite{augment0}, FilterAugment \cite{augment1}, and Mixup \cite{augment2} in \cref{t:augmentimpact1}. SpecAugment applies time and frequency masking as well as time warping. FilterAugment introduces different frequency filters to the spectrograms and Mixup can be described as a weighted combination of two training samples.
We hypothesise that applying masking techniques or time stretching can help by introducing variance but can also have adverse effects, such as loss of information (e.\,g., by masking a step), resulting in worse performance compared to using no augmentation.
We then tested Mixup with four alpha values ($\alpha \in \{0.1,0.3,0.7,1.0\}$) based on a random permutation of batch indices. 
Overall, none of the augmentations led to better predictions than using no augmentations, although SpecAugment and FilterAugment have shown positive results on some models.
Mixup initially seemed promising, as it achieved significant success in related tasks, such as in PANNs \cite{panns}. 
However, these related works, were often based on solving multi-class classification problems, including more prolonged and distinct waveform profiles; the short-term events we are considering here might not be amenable to improvement via the same augmentations. 


\begin{table}[!t]
\begin{center}
\caption{Augmentations on WvLmCnn14 \cite{panns}}\label{t:augmentimpact1}
\scalebox{.9}{
\begin{tabular}{ccc}
\toprule
         PCC      &MAE     &Augmentation         \\
\midrule
         \textbf{0.528}   &\textbf{0.847}   &None\\
         0.472   &0.891   &FilterAug \cite{augment1}\\
         0.478   &0.941   &Mixup($\alpha=0.3$) \cite{augment2}\\
         0.481   &0.983   &SpecAug \cite{augment0}\\
\bottomrule
\end{tabular}}
\end{center}
\end{table}


\subsubsection{Transfer Learning}

Previous work on PANNs \cite{panns}, PSLA \cite{psla}, AST \cite{ast}, and PaSST \cite{passt} has shown that ImageNet pre-training is often more beneficial than pre-training on AudioSet \cite{audioset} or FSD50k \cite{fsd50k}.
We proceed to explore this finding here using PSLA, since the authors provided a wide range of different checkpoints.
Our results are shown in \cref{t:pretrainimpact} and verify the previous observation that ImageNet pre-training is most beneficial.
Surprisingly, random \linebreak weights beat those trained on FSD50K in terms of \ac{MAE}.
This is an instance of a ``negative transfer'' which has been previously observed for other task combinations~\cite{Triantafyllopoulos21-TRO} and is worth a closer examination in future work.


\begin{table}[!t]
\begin{center}
\caption{Different pretrained weights on PSLA \cite{psla}}\label{t:pretrainimpact}
\scalebox{.9}{
\begin{tabular}{ccc}
\toprule
        PCC         &MAE     &Pretrain-weights\\
       \midrule 
        \textbf{0.534}    &\textbf{0.814}   &ImageNet \\
        0.459   &0.875   &AudioSet \cite{audioset} \\
        0.443   &0.879   &Random \\
        0.320   &0.911   &FSD50K \cite{fsd50k} \\
\bottomrule  
\end{tabular}}
\end{center}
\end{table}

\subsubsection{Windowing}

The choice of different window sizes also has a notable impact on the prediction results.
\Cref{t:restimpact} shows 6 different starting windows, based on the first step occurring at 0.0 seconds.
Approach 1, the first three rows, show the already discussed, evenly segmented windows of 5, 10, and 20 seconds.
Approach 2 included starting and closing each window on the basis of step occurrences. 
The window length was still based on t $ \in $ \{5,10,20\} seconds, but usually shorter, stopping and consecutively starting at the last step before the next time interval.
This evaluation is performed for WvLmCnn14, which showed the best balance of performance and total training runtime (measured as total time spent on training; omitted here for brevity).

We note that MAEs are not comparable across the different windows, but rather needs to be multiplied by 4 for a window of 5 seconds and 2 for the 10-second one.
We thus compute a calibrated version of the MAE (cMAE) with that transformation.
Overall, the window size of 5 seconds yielded better results than 10 or 20 -- even when adjusting for calibration.
Furthermore, approach 2 of windowing saw an increase in prediction performance by almost twice as much compared to the Baseline, and PCC values of over 0.746 for  WvLmCnn14.
We note that this approach presupposes ground truth step estimations even for testing, which makes it unrealistic for real-life applications, but shows the promise of combining step counting with segmentation of individual steps, something that can be explored in future work.
We assume this data leakage is what led to the substantial improvement in performance mentioned above.

\begin{table}[t]
\begin{center}
\caption{
MAE and cMAE for different windowing strategies.
}\label{t:restimpact}
\scalebox{.9}{
\begin{tabular}{cccc}
        \hline 
        Threshold (t) is \textit{time}   &  MAE      & cMAE & Baseline   \\
        \hline 
        0.0 - 5.0s	            & 0.847 & \textbf{3.388}     & 1.100          \\
        0.0 - 10.0s 	        & 2.025 & 4.050     & 2.125          \\
        0.0 - 20.0s	            & 4.047 & 4.047         & 4.089          \\
        \hline
        t is \textit{last step} before time                 &MAE & cMAE        &Baseline   \\
        \hline
        0.0 - 4.749s $\leq$ 5.0s	        & 0.506 & \textbf{2.024}      & 1.034          \\
        0.0 - 9.818s $\leq$ 10.0s	        & 1.577 & 3.154      & 1.991          \\
        0.0 - 19.68s $\leq$ 20.0s	        & 4.047 & 4.047    & 4.076          \\
	\hline 
\end{tabular}}
\end{center}
\end{table}


\section{Conclusion}
\label{sec:conclusion}

Using neural networks, we explored the ability to count the number of steps in discretely timed windows of audio by regressing them to a scalar number. 
The experiments included 
a variety of neural networks for each core architecture class, including model weights pre-trained on different data sets. 
Our results show that neural networks are able to reduce the \acl{MAE} of the set of predicted steps (for each window) to the set of actual steps in a way that indicates a meaningful learning behaviour. 
Relatively high values of the \acl{PCC} further suggest a moderate linear correlation for the distribution of predictions compared to the ground truth. 

Our results thus demonstrate the feasibility of estimating the speed of runners by using audio data as a potential alternative to other sensors, which has the potential to lower the barrier of entry for newcomers to the sport.
An intriguing avenue of future research is to recast speed estimation as a sound event detection problem which attempts to identify the individual steps and, optionally, the tibia accelerations that accompany them.
Furthermore, the model weights trained on step detection could possibly improve the results for similar audio-based running-related tasks such as fatigue prediction \cite{kirun0} or surface-type classification \cite{kirun1, kirun2} given their connection to step sounds.


\section*{Acknowledgments}
Results in this study have been partially included in Philipp Wagner's Bachelor Thesis.


\bibliographystyle{ieeetr}
\bibliography{ref}


\end{document}